\documentclass[preprint,12pt]{elsarticle}
\usepackage{setspace}
\usepackage[]{natbib}
\usepackage{graphicx,amssymb,amsmath,times}
\usepackage{subfigure}
\usepackage{lscape}
\journal{New Astronomy}
\begin{document}
\begin{frontmatter}
\title{Intrinsic VHE Gamma-ray spectra of Blazars as a probe for Extragalactic Background Light}
\author{K.K.Singh\corref{corr}}
\ead{kksastro@barc.gov.in}
\author{S.Sahayanathan} \author{A.K.Tickoo} \author{N.Bhatt} 
\address {Astrophysical  Sciences  Division, Bhabha Atomic Research Centre. \\
Mumbai - 400 085, India.}

%--------------------------------Abstract---------------------------------------------

\begin{abstract}
Very high energy (VHE) $\gamma$-rays above 10$'$s of GeV energy, emitted
from distant blazars, are attenuated by photons from the extragalactic
background light (EBL). Unfortunately, neither the EBL nor the
intrinsic blazar spectrum is accurately known to derive one quantity
from the other. In this work we use a homogeneous one zone model involving
synchrotron, synchrotron self Compton (SSC) and external Compton (EC)
emission mechanisms to estimate the intrinsic VHE spectra of blazars.
The model is applied on three VHE blazars, namely PKS2155-304, RGB J0710+591 and 3C279,
for which simultaneous multi-wavelength data are available from various
observations. The predicted values of the intrinsic VHE fluxes are
then compared  with the observations by imaging atmospheric Cherenkov
telescopes to determine the optical depth of VHE $\gamma$-rays. On
comparing these optical depth values with those predicted by four
different EBL models, we observe a somewhat pronounced systematic
deviation for PKS2155-304 and 3C279 at higher energies, especially for
the EBL model proposed by Finke et al.(2010). We attribute this deviation to be an outcome of
either the failure of the extrapolation of blazar SED to VHE energies
and/or due to various assumptions buried in the EBL models.
\end{abstract}

%----------------------------------Keywords--------------------------------------------
\begin{keyword}
Extragalactic Background Light: Opacity, VHE Blazars :PKS 2155-304, RGB J0710+591, 3C 279
\end{keyword}

\end{frontmatter}

%------------------------------------Section:1 Introduction----------------------------

\section{Introduction}
The diffuse cosmic radiation from ultraviolet (UV) to infrared (IR) is commonly 
referred to as extragalactic background light (EBL). EBL is the second dominant
component of cosmic radiation after cosmic microwave background.
It is the contribution of the radiative energy released during the formation of stars and 
galaxies by gravitational and nuclear processes. The spectral energy distribution (SED) 
of EBL is bimodal with high energy component peaking at $\lambda$ $\sim $ 1$\mu$m 
and the second component peaking at $\lambda$ $\sim $100$\mu$m \citep{MGH01,RAB02}. 
Significant part of the high energy component consists of radiation released by 
stars and galaxies whereas the low energy component is due to re-emission of the 
absorbed high energy radiation by dust. Direct determination of the EBL can be done 
by measuring the diffuse emission after subtracting the foreground sources. 
However the contamination by foreground zodiacal light and galactic light 
introduces large uncertainties in such measurements \citet{RAB07,KM03}.
The presence of EBL photons at UV-optical and IR energies can absorb the VHE photons in GeV--TeV
regime emitted from distant sources through pair production \cite{RJG66}. 
Hence estimation of EBL intensity plays an important role 
in knowing the intrinsic VHE spectra of distant sources \cite{FA06,DM07,JA08}. \\

Blazars are the most common VHE gamma ray sources, located at various redshifts,
whose intrinsic VHE spectrum gets modified by EBL absorption.
They are the class of active galactic nuclei (AGN) for which the relativistic jet is aligned close
to the line of sight \cite{Urry95}. Due to relativistic motion, the emission from the jet is Doppler
boosted and dominates the entire SED. Blazars are further classified into flat 
spectrum radio quasars (FSRQ)
and BL Lacs based on the presence/absence of emission line features in their spectra. 
Their SED extends from radio to gamma rays and are bimodal in nature. The first
hump is generally attributed to synchrotron radiation from a non-thermal distribution
of electrons cooling in a magnetic field. Whereas the second high energy component, often extending
up to VHE energies, is believed to be inverse Compton emission by the same electron
distribution, scattering off soft target photons. The soft target photons can be either
synchrotron photons itself (SSC) or the photons external to the jet (EC) \cite{MG85,BS87,GM89,DSM92}.
The SED of BL Lacs are readily explained by considering synchrotron and SSC emission processes
alone whereas, for FSRQ one needs to consider EC also \cite{HBA01}. Moreover detection of 
hard VHE spectra of FSRQ demands that the plausible target photons for the inverse Compton scattering 
may be the IR photons from the obscurring dusty torus suggested by the unification theory of
AGN \cite{Urry95,GT09,SG12} .\\

Presently, the number of blazars detected at VHE energies have increased considerably
\footnote{{http://tevcat.uchicago.edu/}} with the advent of high sensitivity ground based 
Imaging Atmospheric Cherenkov telescopes (IACTs) like \emph{MAGIC} (Major Atmospheric Gamma-ray 
Imaging Camera)\footnote{{http://magic.mppmu.mpg.de/}}, \emph{VERITAS} (Very Energetic Radiation 
Imaging Telescope Array System)\footnote{{http://veritas.sao.arizona.edu/}}
 and \emph{HESS} (High Energy Stereoscopic System)\footnote{{http://www.mpi-hd.mpg.de/hfm/HESS/}}.
Using the observed VHE spectrum, EBL intensity can be constrained 
provided the intrinsic spectrum can be anticipated through various other assumptions. 
Stanev \& Franschini used the TeV spectrum of MRK 501 during a flare observed by
\emph{HEGRA} IACT to estimate EBL intensity \cite{TS98} . 
In their approach, the intrinsic TeV spectra of MRK 501 is assumed to be a power-law
and the spectral shape of EBL is chosen to reflect the average galactic spectrum.
The normalisation of the EBL spectrum and the power-law index of the intrinsic spectrum
is set as a free parameter and the fitting is performed to obtain the observed TeV spectrum.
Aharonian et al (2006) estimated
an upper limit of the EBL intensity from the observed VHE spectra of the blazars H 2536-309 
and 1ES 1101-232. They considered the fact that the hardest particle spectrum attained
through shock acceleration models cannot produce intrinsic VHE spectra flatter than 0.5.
Using this as a hard limit they provided an upper limit to the EBL spectrum.
However an intrinsic VHE spectra flatter than 0.5 can still be produced if one considers
the interaction of the gamma-ray with the surrounding photon distribution or if the  
underlying particle distribution has a relatively high minimum energy \cite{KK06,Aha08}.
One can also exploit the X-ray-TeV correlation, often observed during blazar
flares, to estimate the EBL intensity. This correlation suggests the emissions at these
energies are produced by a common electron population. With the information available
from X-ray observation, one can constrain the basic particle distribution which in turn 
can be used to predict the intrinsic VHE spectrum and the EBL intensity successively \cite{PSC99}. 
Mankuzhiyil et al \cite{NM10} used the broadband spectrum of the BL Lac object, PKS 2155-304, to
estimate the intrinsic VHE spectrum considering synchrotron and SSC emission processes
from a broken power-law distribution of electrons. With this knowledge, the optical depth due to pair production 
with EBL photons is estimated from the observed VHE spectra. This optical depth is
then compared with the one obtained from different EBL models for the given source
redshift. Recently Ackerman et al \cite{MA12} used ~150 BL Lac objects detected by {\emph Fermi} 
satellite at gamma ray energies to predict the EBL spectrum. They reported the
presence of an absorption feature in the GeV spectrum of these sources and used this
feature to estimate the EBL intensity at optical to UV wavelengths.\\

In the present work, we choose three well studied VHE blazars located at different redshifts, 
PKS 2155-304 (z=0.116), RGB J0710+591 (z=0.125) and 3C 279 (z=0.536) for which 
simultaneous or quasi-simultaneous multi band data is available. Out of these three blazars,
 PKS 2155-304 and RGB J0710+591 are BL Lac objects and their intrinsic SEDs are obtained 
considering synchrotron and SSC processes. 
Whereas for 3C 279, being a FSRQ, we include the external Compton scattering of IR photons from 
dusty torus to obtain the intrinsic VHE spectra. In the next section we briefly describe the EBL 
models chosen for our present study and in Section 3, we discuss the process of EBL absorption and
estimation of optical depth of VHE $\gamma$-rays. In Section 4, we explain the model considered to
obtain the intrinsic VHE spectra of blazars. In Section 5, we present 
the results obtained and discuss the highlights of this study. A cosmology with $\Omega_m = 0.3$, 
$\Omega_\Lambda = 0.7$ and $H_0 = 70\;km\;s^{-1}\;Mpc^{-1}$ is adapted 
in this work.

%-------------------------Section:2 EBL Models----------------------------------------------

\section{EBL Models}
Bright contamination of the foreground radiation hampers the direct measurement of EBL
spectrum and hence the EBL intensity is generally estimated empirically. The empirical models 
involve total emission from the galaxies and stars integrated over the redshift to estimate the 
EBL intensity. The spectral properties of the galaxies and stars can be obtained by 
evolving them from cosmological initial conditions or backward extrapolation of the local 
galaxy population belonging to the present epoch. The former is commonly referred as 
forward evolution (FE) models\cite{JRP01,JRP05} and the latter as backward evolution 
(BE) models \cite{SMS06}. The evolution in BE models can be introduced as pure luminosity 
evolution or pure density evolution. The evolution is introduced as $(1+z)^s$ where $z$ 
is the redshift and the index $s$ governs the evolution. In general the parameter $s$ 
can be a function of $z$ though it is often assumed to be constant over a range of 
redshift \cite{MS98}. In FE models, the EBL intensity is estimated while evolving 
the galaxies from initial epoch of star formation. These models use numerical codes to study the 
evolution of stellar population and the resultant SED of the galaxy as a function 
of $z$. More advanced models use semi-analytical approach to include the formation 
of galaxies and their interactions \cite{KTM02,KTM04}. These models obtain the EBL intensity 
at various redshifts by adjusting the model parameters to reproduce the observed 
properties of the local universe. However due to various assumptions and uncertainties 
in estimating the parameters, the EBL intensity suggested by various models differs considerably. 
Here we briefly describe some of the commonly used EBL models and obtain the en route optical depth of 
VHE photons due to them as discussed in Section 3.

\subsection{Franschini, Rodighiero \& Vaccari (2008)}
Franschini et al 2008 \cite{AF08} used an empirical BE model to estimate EBL contribution using 
extensive data set available from  \emph{Spitzer Space Telescope IR camera} 
and ground based telescopes. They assumed the galaxy population is dominated by 
the evolution of spheroidal galaxies, spiral galaxies and merger systems. 
In case of spirals, the comoving number density remains constant after the formation 
at a given redshift but their luminosity evolves along with the evolution of their 
stellar content. For merger systems, they considered the evolution of both, luminosity 
and galaxy number density to be consistent with the observations. Spheroidal galaxies
are assumed to be formed at moderate redshifts and their evolution is achieved by
dividing them into seven sub-populations forming at various redshifts. Assuming 
the mass and luminosity functions do not vary among the sub-populations, they 
set their normalisations to reproduce the local luminosity function. 
The cosmological observables at near IR wavelength are then obtained by adding the 
contribution from these three galaxy classes. A synthetic spectral energy distribution
is then used to extrapolate these observables to shorter wavelength. At longer IR wavelength,
the photons are produced predominantly by the dust thermal emission present in the 
galaxy interstellar medium. Most luminous and massive galaxies are characterized by 
intense star formation and the emitted radiation is effectively reprocessed in to 
IR by dust rich medium. Hence they are bright in IR and their evolution is  modelled 
similar to near IR wavelength.

\subsection{Gilmore et al. (2009)}
Gilmore et al.2009 \cite{RCG09} used semi analytical approach to estimate the EBL by evolving
the radiation released by galaxies and quasars starting from 
cosmological initial conditions. The UV luminosity density from galaxies are predicted 
using semi analytical model of galaxy evolution. Their model also includes the formation
of super massive black holes leading to quasars and AGN. However the  properties of quasars and
AGN predicted by their model is not yet tested with the observations and hence their 
contribution to UV background is added empirically. The emissivities due to galaxies and 
quasars are then integrated over redshifts to predict the evolving UV background of the 
Universe. The parameters of the semi analytical model are constrained to reproduce the 
local galaxy observations. Along with this evolutionary model they also employ a radiative
transfer code to estimate the absorption and remission of ionizing UV radiation by 
inter galactic medium. The re-emission of radiation in IR by the dust is modelled using the 
formalism described by Devriendt \& Guiderdoni (2000) \cite{JEG2000}. In their formalism, the dust and stars 
homogeneously fill the galaxies which are assumed as  oblate ellipsoids. The contribution
of different components of the dust is set to reproduce the local IR background.

\subsection{Finke, Razzaque \& Dermer (2010)}
Finke et al. 2010 \cite{JDF10} predict the intensity of EBL directly from the stellar radiation 
and the reprocessed radiation by dust in the inter stellar medium without involving 
complex semi analytic models. At shorter wavelengths, EBL is dominated by stellar 
emission which is treated as a blackbody. The stellar properties belonging to main
sequence and off main sequence stars are obtained from approximate formulae given by
Eggleton et al. (1989) \cite{Egg89}. Finally their contribution to EBL is estimated through 
initial mass function, stellar formation rate density and the fraction of photons 
escape from being absorbed by the dust in the interstellar medium. At larger 
wavelengths, the emission is dominated by dust which is approximated as a 
combination of three blackbodies. A warm component at temperature $40\;K$ 
representing large dust grains found in and around star forming regions, a hot
component at temperature $70\;K$ representing small dust grains present in the
disk of the spiral galaxies and emission from polycyclic aromatic hydrocarbons
which is assumed as a blackbody at temperature $450\;K$. The resultant IR
emission is calculated self consistently by equating the luminosity density from 
dust emission with the luminosity density from starlight absorbed by the dust.

\subsection{Kneiske \& Dole (2010)}
EBL predicted by Kneiske \& Dole 2010 \cite{KTM10} considers the evolution of star light
using a simple stellar population model for different stellar masses.
For a given initial mass function the evolution is mainly governed by the 
stellar formation rate density. The star forming regions are divided
into two types namely ``optical star forming region'' with low dust
extinction representing luminous infrared galaxies and ``infrared 
star forming region'' with high dust extinction representing ultra
luminous infrared galaxies. The resultant SED produced by a population 
of stars is generated using a spectral synthesis model along with 
dust absorption/reemission model. The IR emission from the dust is 
again approximated as a combination of three black body spectra at
different temperatures. The stellar formation rate density and dust
parameters and then adjusted to reproduce the observed lower limit
of EBL obtained through integrating the number counts of galaxy from 
deep sky survey and completeness correction. \\
The spectral energy density of EBL photons predicted by four EBL models as described above 
is depicted in fig 1. 

%---------------------------------Section:3-EBL Absorption----------------------------------
\section{EBL Absorption}
VHE $\gamma$-rays enroute from source to observer suffer absorption by interaction with the low energy 
EBL photons via pair-production mechanism \cite{RJG67}:  
\begin{equation}
 \gamma_{VHE} + \gamma_{EBL} \rightarrow e^{-} + e^{+}
\end{equation}
The above process is kinematically allowed provided following condition is satisfied by the energies of
two photons,
\begin{equation}
 E \varepsilon (1-cos\phi)=2m_{e}^{2}c^{4}
\end{equation}
where $E$ and $\varepsilon$ are the energies of VHE $\gamma$-rays and EBL photons respectively,
$\phi$ is the scattering angle between momenta of two photons in lab frame and {m$_{e}$} is the 
rest mass of electron. The total cross section for the pair creation process depends 
on the energy of two photons and the angle between them and is given by \cite{BW34}:
\begin{equation}
	\sigma_{\gamma\gamma}(E_{\gamma},\varepsilon,\phi)=\frac{\pi r_{e}^{2}}{2}
 	(1-\beta^{2})[(3-\beta^{4})ln\frac{1+\beta}{1-\beta}-2\beta(2-\beta^{2})]
\end{equation}
where $r_{e}$ is the classical electron radius and $\beta$ represents Lorentz factor  in  
units of velocity of $e^{-}$ or $e^{+}$. In the center of mass frame:
\begin{equation}
	\beta=\biggl[1-\frac{2m_{e}^{2}c^{4}}{E\varepsilon(1-cos\phi)}\biggr]^{1/2}
\end{equation}
The optical depth, $\tau$ encountered by VHE $\gamma$-rays of energy $E$ emitted from a source at 
redshift $z_{s}$ due to EBL absorption is  computed by convolving the EBL photon number density
$n_{EBL}$($\varepsilon$,z) with pair production cross section $\sigma_{\gamma\gamma}(E,\varepsilon,\phi)$.
For cosmological applications, note that $E$ and $\varepsilon$ change along the line of sight in 
proportion to $(1+z)$ due to the expansion of Universe. The three fold integral over the redshift ($z$), 
scattering angle ($\phi$) and the energy of EBL photons ($\varepsilon$) is given by,
\begin{equation}\label{eq:opacity}
	\tau(E,z_{s})=\int\limits_{0}^{z_{s}}\left(\frac{dl}{dz}\right)dz
	              \int\limits_{0}^{\pi}\left(\frac{1-cos\phi}{2}\right)sin\phi d\phi
                      \int\limits_{\varepsilon_{th}}^{\infty}n_{EBL}(\varepsilon,z)
		      \sigma_{\gamma\gamma}(E,\varepsilon,\phi)d\varepsilon
\end{equation}
where the distance travelled by a VHE photon from source to obsrerver in $\Lambda$CDM cosmology is 
expressed as,
\begin{equation}
	\frac{dl}{dz}=\frac{c}{H_{0}}\frac{1}{(1+z)\sqrt{\Omega_{\Lambda}+\Omega_{m}(1+z)^{3}}}
\end{equation}
and threshold energy of background photons for pair production is given by,
\begin{equation}
	\varepsilon_{th}(E,\phi,z)=\frac{2m_{e}^{2}c^{4}}{E(1+z)^{2}(1-cos\phi)}
\end{equation} 
The optical depth of VHE $\gamma$-rays from a given source is estiamted using equation 
(\ref{eq:opacity}) corresponding to different EBL models for $n_{EBL}$($\varepsilon$,z) as described
in Section 2. The intrinsic VHE $\gamma$-ray flux $\left(\frac{dN_{\gamma}}{dE}\right)_{int}$ 
emitted from the source is modified due to EBL absorption and the observed flux 
$\left(\frac{dN_{\gamma}}{dE}\right)_{obs}$ is related to intrisic flux as,
\begin{equation}\label{eq:intvhe}
	\left(\frac{dN_{\gamma}}{dE}\right)_{obs}=\left(\frac{dN_{\gamma}}{dE}\right)_{int} .e^{-\tau(E,z_{s})}
\end{equation}

%--------------------------------Section:4-Blazar Spectrum-----------------
\section{Blazar Spectrum}
We model the intrinsic spectrum of blazars extending from radio to VHE energies as a result of
synchrotron and inverse Compton processes. The emission is assumed to originate from a spherical blob
of radius $R$ moving down the jet at relativistic speed with bulk Lorentz factor $\Gamma$. Since
blazar jets are aligned close to the line of sight of the observer, we assume the Doppler factor 
$\delta = [\Gamma (1-\beta_\Gamma cos\theta)]^{-1}\approx\Gamma$. Here $\beta_\Gamma$ is the jet velocity in 
units of velocity of light $c$ and $\theta$ is the jet viewing angle. The emission region is 
populated uniformly with a broken power law electron distribution described by
\begin{equation}
	N(\gamma)d \gamma = K\left[\left(\frac{\gamma}{\gamma_b}\right)^{p1}+\left(\frac{\gamma}{\gamma_b}\right)^{p2}\right]^{-1} 
	d \gamma \quad ;\gamma_{min}< \gamma< \gamma_{max}
\end{equation}
where $K$ is the normalisation, $\gamma_b m_e c^2$ is the break energy with $m_e$ as the 
electron rest mass, $p1$ and $p2$ are the 
power law indices before and after the break energy $\gamma_b m_ec^2$ and $\gamma_{min}m_ec^2$ and 
$\gamma_{max}m_ec^2$ are the minimum and the maximum electron energy of the distribution. 
A broken power law particle distribution can be result of various physical conditions. For example, 
radiative cooling of a simple power law particle distribution can produce a broken power law with 
indices related by $p_{2}=p_{1}+1$. The break energy $\gamma_{b}$ then decides the age of the 
emission region \cite{SS03}. If the radiation is due to synchrotron and/or inverse Compton processes, 
the observed spectral indices will differ by canonical 0.5. However, if the observed spectral indices fail 
to satisfy this condition, the underlying broken power law distribution can be an outcome of complex 
situations probably involving more than one acceleration processes \cite{SS08}. In the present work, the 
indices $p1$ and $p2$ are considered as free parameters resulting from any of the above mentioned mechanisms. 
The synchrotron spectrum is obtained due to cooling of this particle distribution in a 
tangled magnetic field $B_{eq}$. The magnetic field $B_{eq}$ and the relativistic particle 
distribution is assumed to be in equipartition
\begin{equation}\label{eq:partdist}
	U_B = m_e c^2 \int\limits_{\gamma_{min}}^{\gamma_{max}} \gamma N(\gamma)  d\gamma = U_e
\end{equation}
where $U_B = B_{eq}^2/8\pi$ is the magnetic field energy density and $U_e$ is the particle energy
density. The synchrotron emissivity at a given frequency $\nu$ is calculated by convolving the 
electron spectrum $N$($\gamma$) with single particle emissivity P($\nu_{s}$,$\gamma$) averaged over an isotropic 
distribution of pitch angles and is given by \cite{RL79},
\begin{equation}\label{eq:sync}
	j_s(\nu_s)=\frac{1}{4\pi}\int\limits_{\gamma_{min}}^{\gamma_{max}} N(\gamma) P(\nu_s,\gamma) d\gamma
\end{equation}
High energy emission is attributed to inverse Compton emission where soft target 
photons are scattered to high energy by the electron distribution given by $N$($\gamma$).
The target photons can be either synchrotron photons itself (SSC) or the photons external 
to the jet (EC). The inverse Compton emissivity is given by \cite{DS93},
\begin{equation}\label{eq:IC}
	j_{IC}(\varepsilon, \Omega)=m_e c^3 \varepsilon \int\limits_{0}^{\infty} d\varepsilon^\prime 
				    \oint d\Omega^\prime \int\limits_{\gamma_{min}}^{\gamma_{max}} d\gamma
                                    (1-\beta cos\psi)n_{ph}(\varepsilon^\prime,\Omega^\prime) N(\gamma)
				    \sigma(\varepsilon, \varepsilon^\prime, \Omega^\prime)	 
\end{equation}
where $\psi$ is the angle between the incident photon and electron directions,  
$\sigma$($\varepsilon$, $\varepsilon^\prime$, $\Omega^\prime$) is the scattering cross section and 
$n_{ph}$($\varepsilon^\prime$, $\Omega^\prime$) is the target photon distribution in blob frame. 
For an isotropic target photon distribution, $n_{ph}$($\varepsilon^\prime$, $\Omega^\prime$) can 
be replaced by an angle averaged distribution. Since BL Lac objects lack line/thermal emission 
we consider only synchrotron and SSC process to describe their broadband SED. Whereas, for FSRQ 
EC process should also be taken into consideration since line and/or thermal features are 
significant in their SED. Moreover modelling their broadband SED also demand the need of 
EC process to explain the $\gamma$--ray spectrum \cite{HBA01}. Further, FSRQ which are observed 
with a hard VHE spectrum suggest that the scattering process must be in 
Thomson regime. This condition can be achieved if the soft target photons are the IR
photons from the dusty torus proposed by the unification theory \cite{GG08,SG12}. 
For SSC emission, the target photon distribution will be isotropic whereas for EC spectrum 
estimation it is assumed to be anisotropic. Finally, the total flux received by the
observer is obtained considering relativistic and cosmological effects for the case of 
no en route absorption using the relation \cite{MCB84},
\begin{equation}
\label{eq:flux}
F_{tot}(\varepsilon)\approx \frac{\delta^3(1+z)}{d_L^2} V^\prime j^\prime\left(\frac{(1+z)}{\delta}\varepsilon
\right)
\end{equation}
where $\varepsilon$ is the observed energy, $d_L$ the luminosity distance, $z$ the redshift of the source,
$V^\prime$ the volume of the emission region and $j^\prime$ is the emissivity due to
different radiative processes.

%--------------------------Section:5-Results & Discussion-----------------------------------

\section{Results and Discussion}

The main parameters describing the intrinsic broadband blazar SED are the particle
indices $p1$ and $p2$, magnetic field $B$, particle normalisation $K$, break energy of the 
particle distribution $\gamma_bmc^2$, size of the emission region $R$ and bulk Lorentz factor 
$\Gamma$. In case of FSRQ, we have additional parameters describing the external photon field.
Among these $p1$ and $p2$ can be obtained through the observed photon spectral indices. Using 
observed synchrotron and SSC fluxes, synchrotron peak frequency, variability timescale and  
equipartition, rest of the parameters can be constrained. For EC process we assume
the external photon field to be blackbody radiation at temperature $T$ which can be constrained
using observed EC flux. The source parameters are optimized by 
reproducing  the observed broadband SED of blazars excluding the VHE data 
since intrinsic VHE spectra is dependent 
on the EBL density. We apply this procedure for three blazars described below to study the transparency 
of VHE photons. 

%--------------------------------Table-Source Parameters---------------------------------
\begin{table}
\caption{Optimized source parameters from multi wavelength data and blazar modeling.}
\begin{center}
\begin{tabular}{lcccccccccc}
\hline
Sources  &\multicolumn{9}{c}{Parameters} \\
	      &z  &R 	&$\Gamma$ &$\gamma_{b}$	&$U_{e}$ &$p_{1}$ &$p_{2}$ &B 	&T \\
\hline
PKS2155-304   &0.116 &0.79 &26.7 &1.83$\times$$10^{4}$ &2.62 &1.82 &4.16 &0.30 &-\\
RGBJ0710+591  &0.125 &0.45 &26.0 &7.83$\times$$10^{4}$ &1.31 &1.89 &3.33 &0.17 &-\\
3C279	      &0.536 &2.17 &25.4 &1.22$\times$$10^{3}$ &10.90&2.05 &4.63 &0.45 &850\\
\hline
\end{tabular}
\end{center}
\textbf{Notes--}   
Col. 1: Source Name; Col. 2: Source Redshift;\\  
Col. 3: Size of emission region (in units of $10^{-2}$pc); \\
Col. 4: Bulk Lorentz factor;  Col. 5: Break energy of particle distribution (in units \\
of electron rest mass energy m$c^{2}$);  Col. 6: Particle energy density (in units  \\
of $10^{-3}$ergs $cm^{-3}$);  Col. 7\&8: Power law indices of particle distribution \\
before and after the break respectively;  Col. 9: Magnetic field (in units of Gauss); \\
Col.10: Temperature of black body radiation in (K) for EC process.
\end{table}
%-------------------------------------------------------------------------------------------
\subsection{PKS 2155--304}
PKS 2155--304 is a high frequency peaked BL Lac object at redshift z=0.116. Dedicated 
multi-wavelength observations of this object including GeV and TeV observations with 
\textit{Fermi}-LAT and H.E.S.S. are reported by Aharonian et al. (2009) \cite{Aha09}. 
We use this multi-wavelength data from optical to HE $\gamma$--rays to model the SED 
using synchrotron and SSC emission processes. 
The optimized parameters describing the SED of the source are 
given in Table 1. The resultant model curve along with multi wavelength data including 
observed VHE fluxes by H.E.S.S. telescope is shown in fig 2(a). The predicted intrisic VHE 
flux is more than the observed one and we account for this discrepancy as a result of absorption
through pair production with EBL photons. From the ratio of these fluxes, we estimate the optical 
depth of this process and compare it with the one estimated through four EBL models discussed in 
section 2. In fig 2(b) we show the optical depth due to different EBL models and the one expected 
from broadband SED modeling.

%-----------------------------------------------------------------------------

\subsection{RGB J0710+591}
RGB J0710+591 is a high frequency peaked BL Lac (HBL) object at redshift z=0.125.
This source was detected by VERITAS telescope during the 2008-09 observation. 
Following this detection, an extensive multi-wavelength observation from optical to 
VHE $\gamma$--rays was initiated and the results were reported by 
Acciari et al. (2010) \cite{Acc10}. We use this multi-wavelength data from optical to
 HE $\gamma$--rays to model the SED of the source using synchrotron and SSC emission processes. 
The optimized source parameters are given in the Table 1. The observed SED of the source including
VHE observations by VERITAS is shown in fig 3(a) along with model curves. The optical depth 
estimated from the ratio of observed and intrinsic VHE fluxes is plotted in fig 3(b) along with 
the one estimated for different EBL models. The large errors in optical depth values, especially 
at high energies, are due to the uncertainties in the observed VHE spectra.  

%-----------------------------------------------------------------------------

\subsection{3C 279}
3C 279 is an FSRQ at a redshift of 0.536 and the farthest blazar 
detected in VHE till now. Hence, its intrinsic VHE $\gamma$--ray spectrum is modified 
considerably by EBL absorption. We use simultaneous observations from optical to VHE 
$\gamma$--rays during a $\gamma$--ray flare in 2006 to obtain the broadband SED of 
the source reported by Albert et al. (2008). Being an FSRQ, the SED of 3C279 is reproduced by 
considering synchrotron, SSC and EC processes. Due to lack of simultaneous observations at MeV, 
we include VHE flux at 84 GeV in multi wavelength data to obtain the optimized 
set of parameters governing the broadband spectrum. The source parameters are given in the 
Table 1 and the model spectrum of the source along with the multi wavelength data upto VHE 
is given in fig 4(a). The optical depth estimated from the ratio of observed and intrinsic VHE 
fluxes and for different EBL models is plotted in  fig 4(b). \\

For all three blazars, a trend of deviation between the optical depth estimated through SED modeling 
from the one obtained from EBL models is observed. For two BL Lacs PKS 2155--304 and RGB J0710+591, 
which are closer compared to the FSRQ 3C 279, we have VHE information upto $\sim$3 TeV. For these two 
sources the difference between optical depths increases with the increase in energy, though for 
RGB J0710+591 we have large uncertainties. Moreover the largest deviation at 3.2 TeV in case of 
PKS 2155--304 and 3.5 TeV in case of RGB J0710+591 is obtained when the EBL model by 
Finke et al. (2010) is used. A closer examination suggests that at low energies the estimated 
optical depth is larger whereas the trend reverses as one moves to higher energies with maximum 
deviation at $\sim$3 TeV. Since these two sources are located at almost similar redshift, the 
evolutionary effects of EBL will not be prominent between them. Hence two possible interpretations 
can be made to understand this deviation. In the first case, if we assume the optical depth predicted 
by SED modeling to be the correct description of EBL intensity, then the EBL models under predict the 
EBL intensity at higher IR frequencies (corresponding to lower VHE photon energies) giving rise to less 
optical depth at lower VHE band. Whereas at higher VHE regime, the EBL models over predict the 
low energy EBL spectrum in IR band giving rise to large optical depth. In that case, the 
assumption made in estimation of IR component in these EBL models through absorption and reemission
of radiation and cosmological initial conditions may give rise to such discrepancies. On the other 
hand, if we assume that the optical depths predicted by these EBL models are correct, then the 
extrapolation of blazar SED using simple emission models may be erroneous. It means at VHE, the 
spectrum may not follow the simple power law expected from lower energies. Such a case is 
possible if the particle distribution hardens at high energies. A possible explanation of this
feature can be due to the Maxwellian tail often encountered at high energies of an accelerated 
particle distribution
where the escape time scale is longer than the cooling or acceleration time scales \cite{SO02}.
Presence of such features can be probed only by VHE observations describing the Compton tail of SED.
The synchrotron regime of SED will not reflect this feature because the high energy tail of the same 
is often buried inside the Compton regime. Alternatively, understanding such features demand precise
knowledge of EBL spectrum to obatain the intrinsic VHE spectra.\\

For 3C 279, we have information only upto 475 GeV. Being the farthest source, the evolution of EBL may 
be prominent and hence the deviation of optical depths will reflect the evolutionary history of the 
Universe also. As expected the deviations are larger even at relatively lower VHE since the uncertainties 
regarding the evolution of EBL will also be folded into the uncertainties arising from the EBL spectrum and 
SED emission model. Even for this source we find a trend similar to BL Lacs that the estimated optical depth 
obtained using SED modeling is larger than the one predicted by EBL models at low VHE regime and the trend
reverses as we move to high energies. Hence the two interpretations suggested earlier for BL Lacs will be 
applicable to this source also. However in this case, the evolutionary history of EBL may also be a prominent 
factor causing the deviation in optical depths. \\

Recently, Abramowski et al. (2013) investigated the EBL absorption feature in the VHE spectra of $HESS$ 
detected blazars using maximum likelihood method \cite{ABR13}. They have assumed a smooth $\gamma$-ray 
spectra and estimated the EBL intensity through intrinsic spectral curvature. They used  EBL shape 
proposed by Franceschini et al. (2008) and obtained the normalisation through fitting procedure.
Yuan  et al. (2012) have also used a Monte Carlo fitting procedure to obtain the model independent 
EBL intensities and intrinsic parameters of blazars \cite{Yuan12}. Due to large uncertainties in 
their predicted EBL intensity, the derived optical depth will not differ considerably from the one 
obtained in the present work. \\

%-----------------------------Section-5-Conclusion------------------------------------------------
\section{Conclusions}

In the present work, a simple emission model considering synchrotron and inverse Compton emission 
mechanisms is used to reproduce the SED of blazars for which simultaneous multi wavelength 
observations are available. The model is applied on three well studied VHE blazars at various 
redshifts. The model is then used to predict the intrinsic VHE spectra of these blazars. 
The source parameters are constrained using multi wavelength data and are in close agreement 
with the values generally considered for blazars. The predicted intrinsic VHE flux by this model 
is used to estimate the optical depth for VHE $\gamma$--rays from the observed flux due to EBL absorption.
We then compare this with optical depth obtained using four commonly used EBL models in VHE regime. \\

The deviation of optical depths is seen more at higher VHE regime with the optical depth predicted by SED 
modeling being lower. However the trend reverses at lower VHE regime though the deviation is not large as 
compared to the high energy end. We interpret this behaviour as an outcome of two possible scenarios. In the first 
case, the discrepancy may be due to under prediction of EBL intensity at higher IR frequencies and over 
prediction at lower IR frequencies by various EBL models. Alternatively, the deviation may also imply 
the failure of extrapolating the SED of blazars to VHE regime using simple emission models. In such case, the 
intrinsic VHE spectrum must be hard suggesting an excess in the high energy tail of the underlying particles 
distribution. With the poor statistics involving only three sources, we are not able to identify the more 
probable interpretation out of these two scenarios. However, we can forsee that identifying the correct EBL 
spectrum will be a potential tool to understand the intrinsic VHE spectrum of blazars which in turn explain 
the underlying particle acceleration mechanism. This can also be used to test different cosmological models 
enabling us to understand the Universe better. At the present epoch we are witnessing the rising number of VHE 
blazars due to various operational atmospheric Cherenkov telescopes, like MAGIC, VERITAS and 
HESS and this can constrain the EBL spectrum substantially. More over, with the help of upcoming 
high sensitivity mega experiment \emph{CTA} (Cherenkov Telescope Array)\footnote{{http://www.cta-observatory.org/}}, 
the number of VHE blazars will increase sharply paving a way for the better understanding about blazars 
and our Universe.

%---------------------------------------------------------------------------------------------
\section*{Acknowledgement}
We would like to thank the anonymous reviewer for his/her suggestions which improved the quality of the paper. KKS thanks 
Dr. N. Mankuzhiyil and Dr. J. S. Perkins for sharing multi wavelength data used in the present study.
%----------------------------References--------------------------------------------

%------------------------------------------------------------------------------------------------
\newpage
%-------------------------------------Fig-1:EBL models-------------------------------------
\begin{figure}
\begin{center}
\includegraphics*[width=0.75\textwidth,angle=-90]{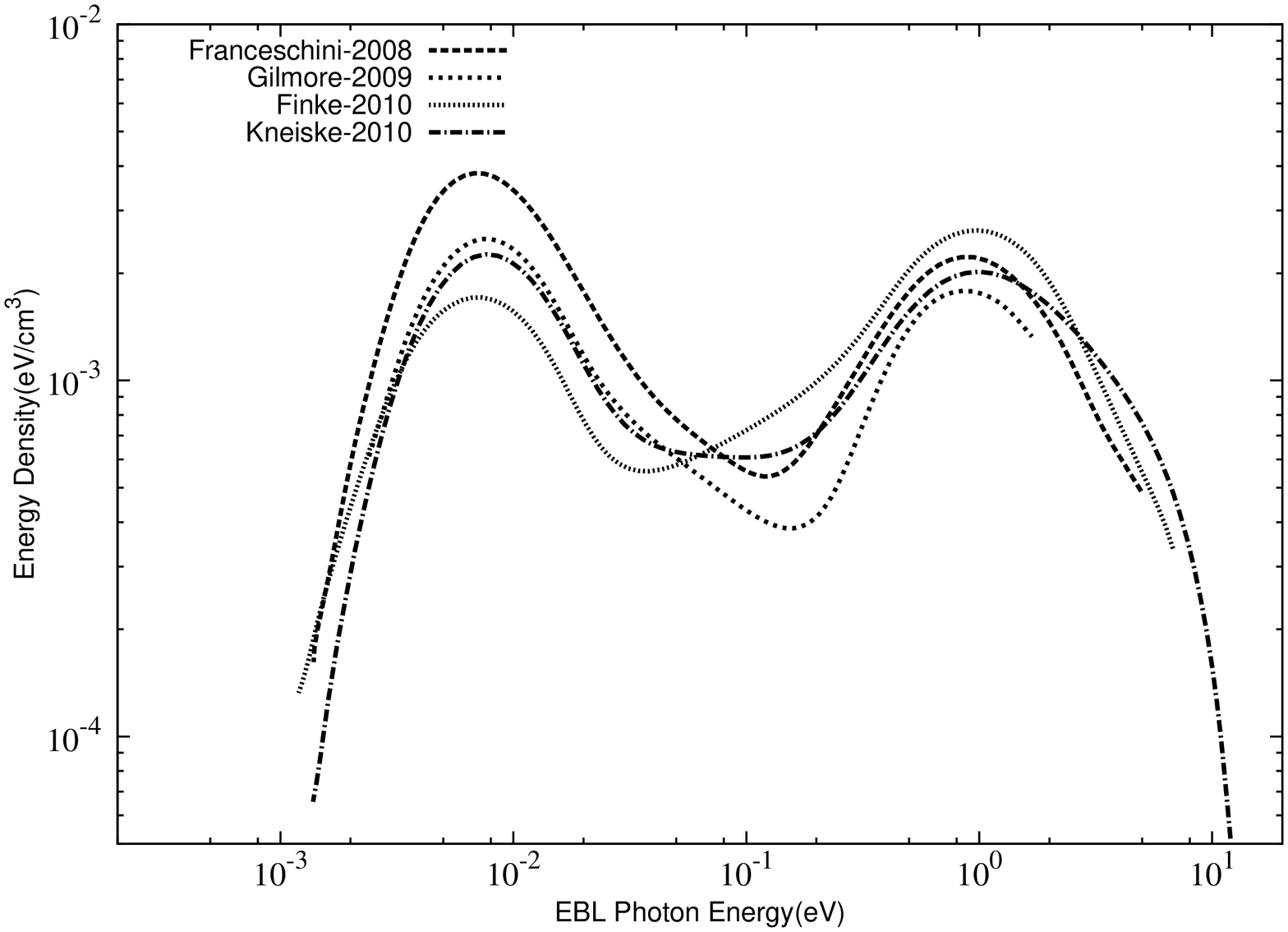}
\caption{Spectral energy distribution of EBL predicted by four different models.}
\end{center}
\end{figure}

%-----------------------------Figure-2-PKS2155-------------------------------
\begin{figure}[h]
\begin{center}
\subfigure[]{\includegraphics*[width=0.7\textwidth,angle=-90]{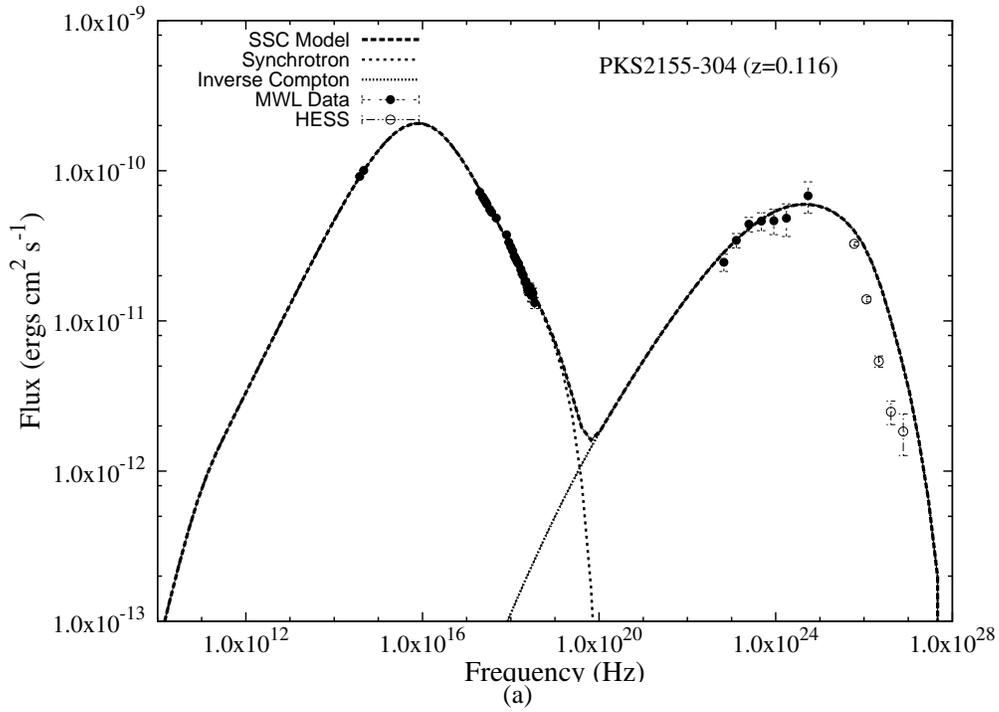}}
\subfigure[]{\includegraphics*[width=0.7\textwidth,angle=-90]{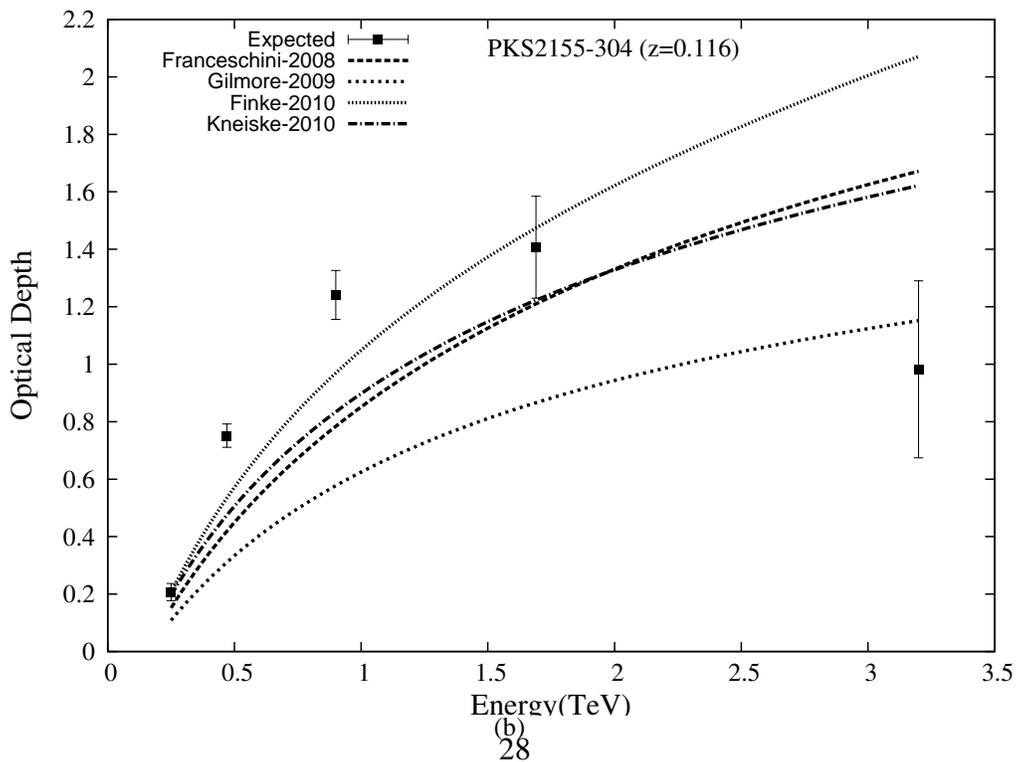}}
\caption{(a) Observed SED of PKS 2155-304 along with model curve. 
 (b) Optical depth predicted through SED modeling and comapred with the one estimated for different EBL 
models.}
\end{center}
\end{figure}

%---------------------------Figure-3-RGBJ0710----------------------------------
\begin{figure}[h]
\begin{center}
\subfigure[]{\includegraphics*[width=0.7\textwidth,angle=-90]{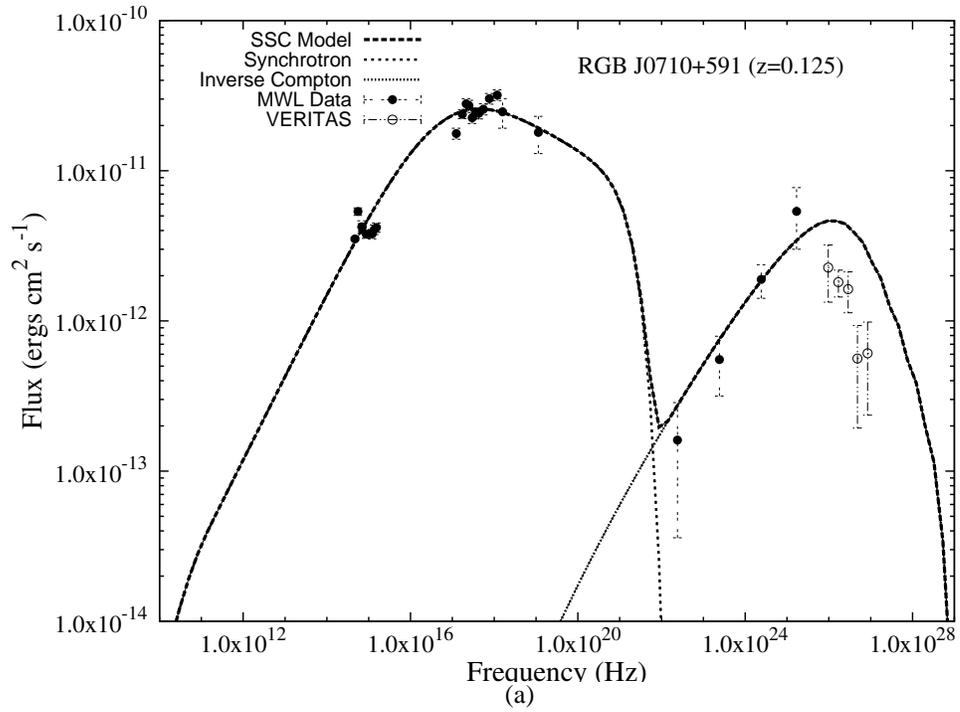}}
\subfigure[]{\includegraphics*[width=0.7\textwidth,angle=-90]{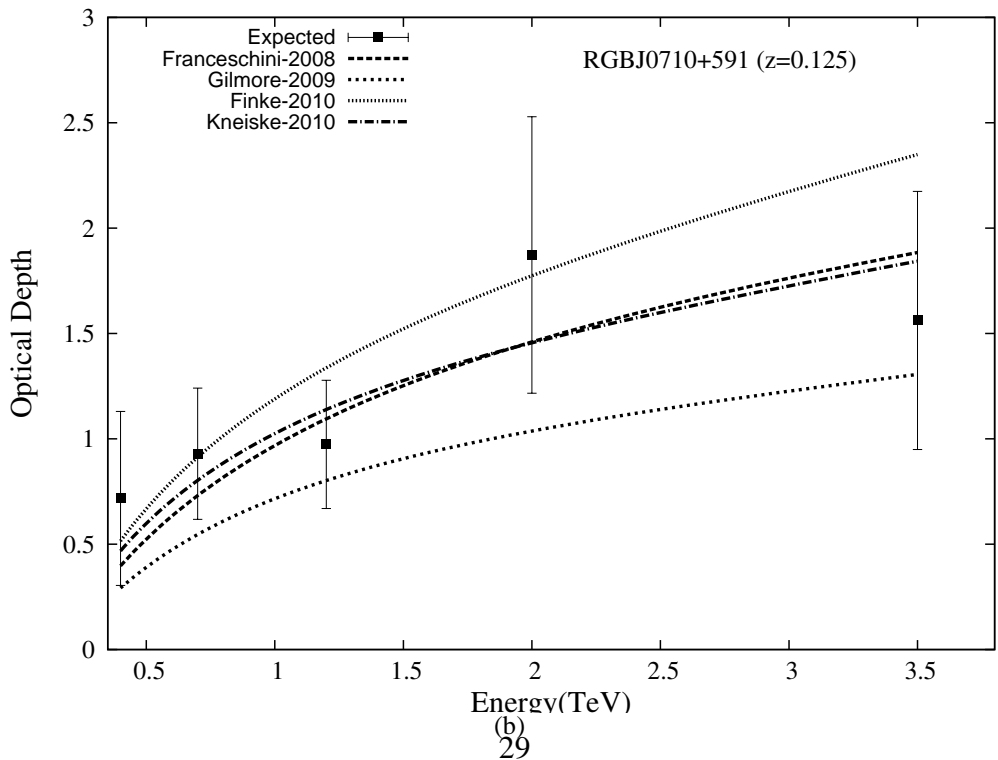}}
\caption{(a) Observed SED of RGB J0710+591 along with model curve.   
(b) Optical depth estimated from SED modeling and different EBL models.}
\end{center}
\end{figure}

%---------------------------Figure-4-3C 279----------------------------------
\begin{figure}[h]
\begin{center}
\subfigure[]{\includegraphics*[width=0.7\textwidth,angle=-90]{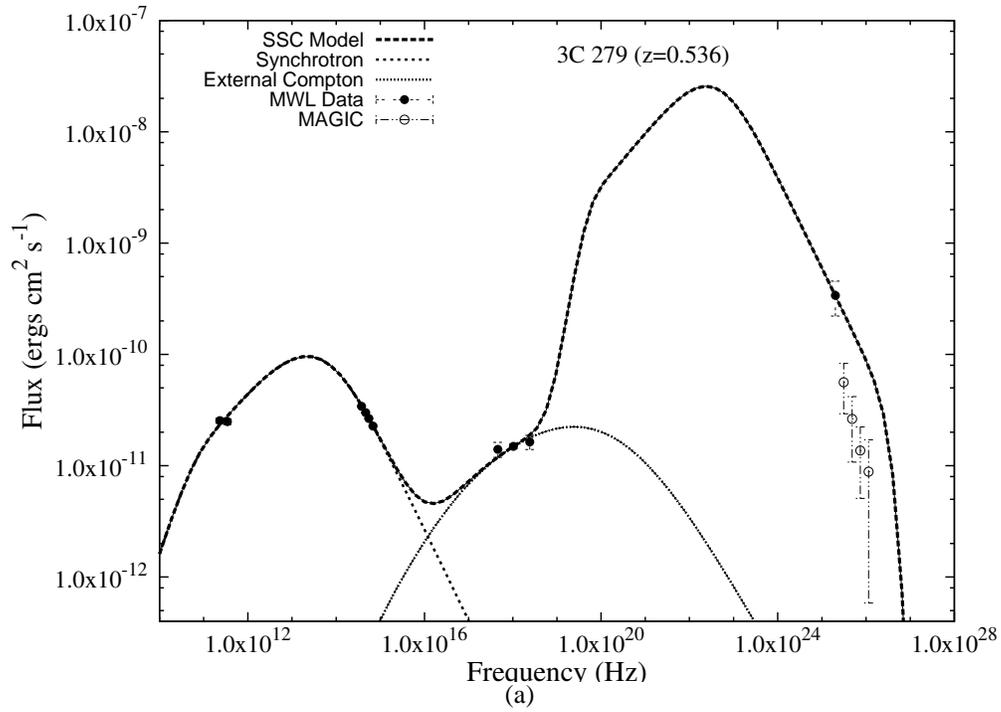}}
\subfigure[]{\includegraphics*[width=0.7\textwidth,angle=-90]{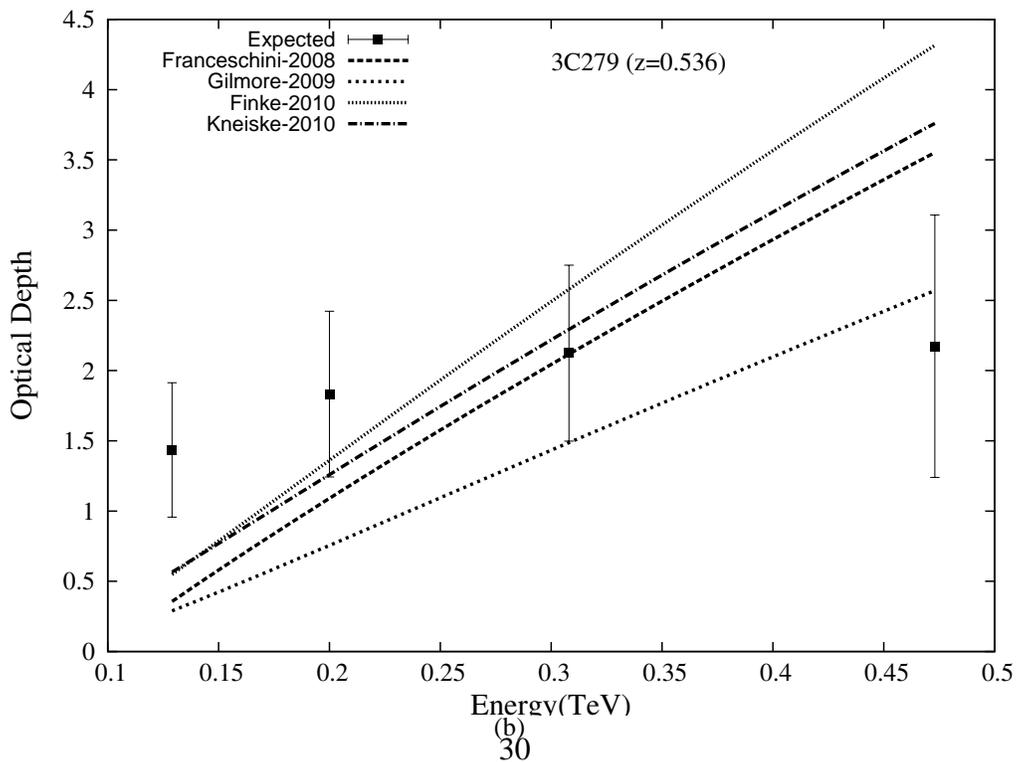}}
\caption{(a) Observed SED of 3C 279 along with model curve. 
(b) Optical depth estimated from SED modeling and different EBL models.}
\end{center}
\end{figure}

\end{document}